# Boosting the efficiency of transient photoluminescence microscopy using cylindrical lenses

*Alvaro J. Magdaleno, Mercy Cutler, Jesse J. Suurmond, Marc Meléndez, Rafael Delgado-Buscalioni, Michael Seitz, Ferry Prins\**

A. Magdaleno, M. Cutler, J. Suurmond, M. Seitz, F. Prins
Condensed Matter Physics Center (IFIMAC) and Department of Condensed Matter Physics, Universidad Autónoma de Madrid, 28049 Madrid, Spain.
E-mail: ferry.prins@uam.es

M. Meléndez, R. Delgado-Buscalioni
Condensed Matter Physics Center (IFIMAC) and Department of Theoretical Condensed Matter Physics, Universidad Autónoma de Madrid, 28049 Madrid, Spain.

**Abstract**

Transient Photoluminescence Microscopy (TPLM) allows for the direct visualization of carrier transport in semiconductor materials with sub nanosecond and few nanometer resolution. The technique is based on measuring changes in the spatial distribution of a diffraction limited population of carriers using spatiotemporal detection of the radiative decay of the carriers. The spatial resolution of TPLM is therefore primarily determined by the signal-to-noise-ratio (SNR). Here we present a method using cylindrical lenses to boost the signal acquisition in TPLM experiments. The resulting asymmetric magnification of the photoluminescence emission of the diffraction limited spot can increase the collection efficiency by more than a factor of 10, significantly reducing acquisition times and further boosting spatial resolution.

# 1. Introduction

Transient Photoluminescence Microscopy (TPLM) has emerged as a powerful technique to directly visualize the displacement of energy carriers in a variety of semiconductor materials, including organic semiconductors,[1] nanocrystal assemblies,[2,3] transition metal dichalcogenides,[4,5] and perovskite semiconductors.[6–9] Through this direct visualization of the carrier motion, detailed information on the spatial dynamics can be obtained.[10]

In its simplest form, TPLM uses a pulsed laser for near-diffraction-limited excitation of an initial population of energy carriers. As the population broadens in time, photoluminescence (PL) emission that results from carrier recombination is collected and projected with large magnification (typically >300×). By raster scanning an avalanche photodiode (APD) through the strongly magnified image plane, a map of the spatial and temporal dynamics of the population can be recorded with sub-ns and few-nm resolution. Raster scanning of an APD is the simplest and most cost-effective implementation of TPLM,[1,2] though alternative detection schemes have been reported as well, including the use of streak cameras[5,7] or time-gated ICCDs.[8] While there are small differences in the quantum efficiency and temporal resolution of these different detectors, the overall performance is comparable.

Using this method, a movie-like representation of the time-dependent broadening of the carrier population is obtained. By determining the change in variance ($\sigma^2(t) - \sigma^2(t = 0)$) of the population for each time-slice, the mean-square-displacement (MSD) of the population can be extracted (MSD = $\sigma^2(t) - \sigma^2(t = 0)$). Importantly, the spatial precision with which the expansion can be determined is set by the signal-to-noise-ratio (SNR) and can be on the order of just 10 nm.[2] Consequently, maximising the SNR is one of the key challenges in TPLM experiments.

Importantly, in many materials, carrier diffusion is isotropic along the different in-plane dimensions. In this case, a line-scan through the centre of the carrier population suffices to obtain the full carrier transport characteristics. The reduced scan area allows for a higher spatial precision within the same acquisition time. At the same time though, a large fraction of the signal is projected outside of the collection area of the line scan and is therefore lost. To further boost the acquisition efficiency of TPLM measurements in isotropic materials, we propose the use of cylindrical lenses to concentrate the majority of the PL signal onto the one-dimensional scan area. We show that this method dramatically improves the SNR for a given acquisition time. While this improvement is significant to all types of materials, it is largest for materials with a large diffusion length where carriers diffuse farther out from the original excitation spot. The use of cylindrical lenses provides a straightforward route toward improved spatial resolution in TPLM measurements.

## 2. Results

Our basic TPLM setup is based on the original reports by Akselrod et al.[1,2] and uses raster scanning of an avalanche photodiode (APD, Micro Photon Devices PDM, 20×20 µm detector size) through the image plane of the carrier population. In short, a pulsed laser diode (PicoQuant LDH-D-C-405, PDL 800-D, λ = 405 nm) is focussed down to a near-diffraction limited excitation spot with a full-width-at-half-maximum (FWHM) of a few hundred nm using a 100× oil immersion objective (Nikon CFI Plan Fluor, N.A. = 1.3). Fluorescence from the material is then collected using the same objective, filtered using a dichroic filter and projected to an intermediate image plane using the standard f = 200 mm tube lens of the microscope to yield a 100× magnification. To generate a higher magnification for TPLM measurements, a set of relay lenses is used. For conventional TPLM, this set consists of a $f$ = 100 mm collimating and a $f$ = 400 mm imaging lens to yield a theoretical total magnification of the image plane of 400× (see **Figure S1**). For example, taking an emission spot of 500 nm, the projected size of the carrier population at this magnification would be as large as 200 µm, which is significantly larger than the APD detector size of 20 µm. The relatively small size of the APD allows for high resolution mapping of the emission profile of the carrier population along the scan direction. Importantly though, as mentioned, a trade-off exists between the achievable resolution in the scan direction and the loss of signal from carriers moving in the orthogonal direction.

To capture signal from carriers moving orthogonally to the scan direction, we implement a detection scheme with cylindrical lenses. In this alternative geometry, we substitute the $f$ = 400 mm spherical lens with a set of two perpendicularly oriented cylindrical lenses to collapse the x-axis of the image plane (see **Figure 1** and a more complete set-up description in **Figure S2**). The first cylindrical lens (LJ1363RM Thorlabs) has the identical focal length to the original spherical lens ($f_1$ = 400 mm) and generates the high total magnification of 400× along the scanning direction (y-axis in Figure 1). The second cylindrical lens has a short focal length ($f_2$ = 20 mm, AYL2520-A Thorlabs) and is used to collapse the image along the orthogonal direction (x-axis in Figure 1), resulting in a total magnification of 20×. With a magnification ratio of 400/20 = 20, the short axis of the asymmetric Image of the carrier population is then reduced to around 10 µm, well within the size of the APD detector and significantly reducing the signal losses due to carriers moving in the orthogonal direction. It should be emphasized that these magnifications are theoretical and can vary in practise due to imperfections in the alignment and in the lens quality.

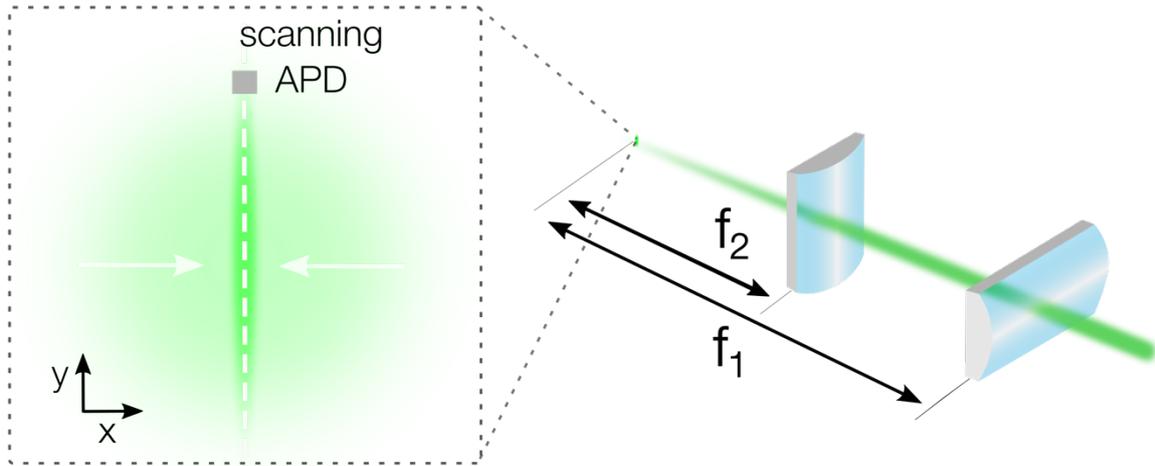

**Figure 1.** Set of cylindrical lenses placed along the light path towards the APD. The lenses highly magnify the PL emission spot along the scanning direction (y-axis) while simultaneously demagnifying it along the orthogonal direction (x-axis). The APD scan of the imaged PL spot is performed along the y-axis (white dashed line).

To test the cylindrical lens setup, we image the PL emission spot of a single crystalline flake of 2D perovskite (($PEA)_2PbI_4$) onto an EMCCD camera (Princeton Instruments, ProEM HS 1024BX3). 2D perovskites exhibit isotropic carrier diffusion in the in-plane directions in the form of excitons.[6,7,11–13] **Figure 2**a shows the image of the PL emission using the traditional spherical lens setup, showing the projection of the symmetric emission spot at large magnification. In comparison, for the image from the cylindrical lens setup shown in **Figure 2**b, the large magnification along the y-axis is maintained but the compression of the image along the x-axis is clearly visible. To compare the light intensity of the cylindrical ($I_{cyl}$) and spherical lens ($I_{sph}$) configurations, in **Figure 2**c we show the intensity profiles along the x-axis normalized to the maximum intensity of $I_{sph}$. For the cylindrical projection the FWHM of the large y-axis magnification (blue solid line) measures 312 µm, while for the x-axis this is reduced to 26 µm, yielding a magnification ratio of around 12. While this is less than the theoretical magnification ratio (400 / 20 = 20), a significant gain in signal collection onto the 20 µm APD detector can be achieved, as indicated by the grey dashed lines in **Figure 2**c. To quantify the signal enhancement, we integrate the intensity of the central 20 µm of the line traces shown in **Figure 2**c. While the spherical lens projects only 3 % of the total emission along the x-axis onto the detector, the cylindrical lens enhances this to 35 %. This is translated into a total integrated intensity enhancement ($I_{cyl}/I_{sph}$) of 10.8.

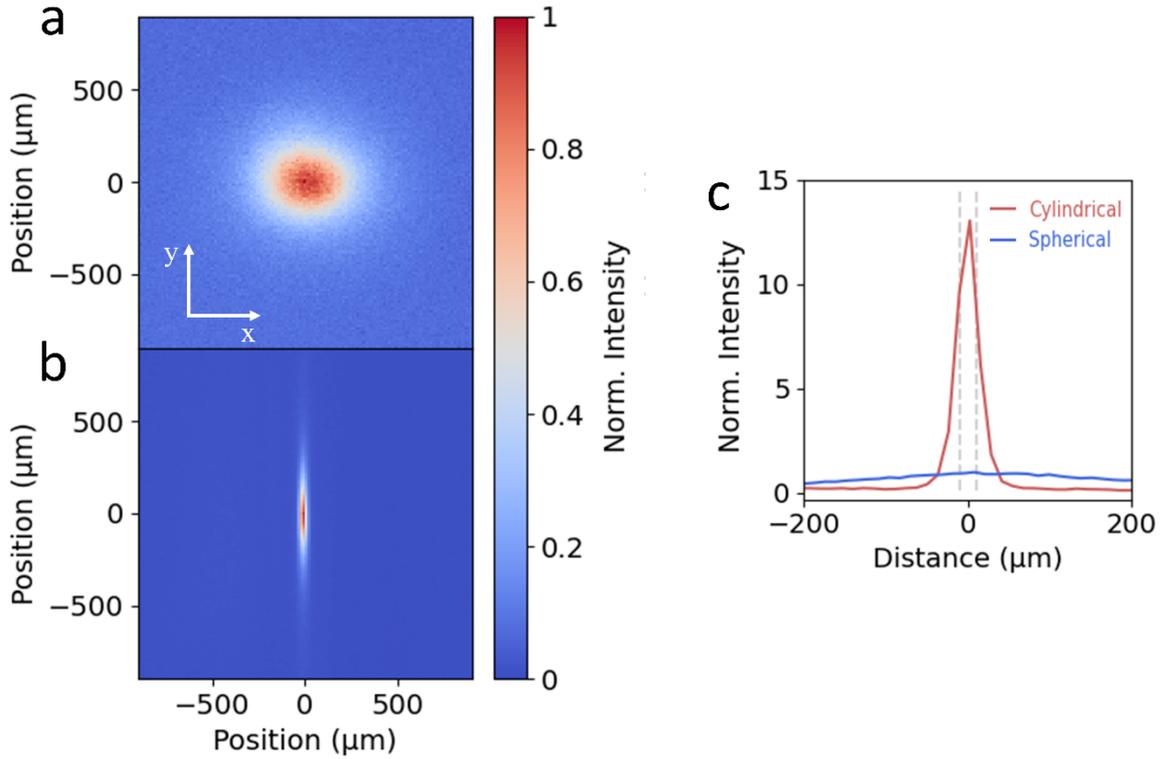

**Figure 2.** (a-b) Visualization of the projection of the emission intensity of the 2D perovskite with the spherical and the cylindrical lenses, respectively. (c) Intensity profiles along the x-axis of the center of the population for the cylindrical (red) and the spherical (blue) lenses. Dashed grey lines denote the APD size (20 µm). Intensities in c are normalized to the maximum intensity detected with the spherical lens.

To test the benefits of the intensity enhancement achieved with the new configuration, we perform TPLM with both set-ups. For this, the image plane is projected onto the APD, which is placed on a motorized x-y stage that allows for the spatial scanning. In a first step, the APD is raster scanned across the image plane to find the center of the emission spot. Then, for both the spherical and cylindrical lens configuration, the actual measurement is performed by scanning along the y-axis while keeping the x-axis fixed. The resulting one-dimensional diffusion maps for the 2D perovskite are presented in **Figure 3**a and b. In these spatiotemporal maps, every time slice is normalized to its maximum intensity, highlighting the broadening of the population as a function of time. Despite equal integration times (4.5 min), laser repetition rate (40 MHz), and excitation fluence (50 nJ/cm$^2$), the SNR of the diffusion map of the cylindrical lens configuration is considerably better as compared to the spherical lens configuration.

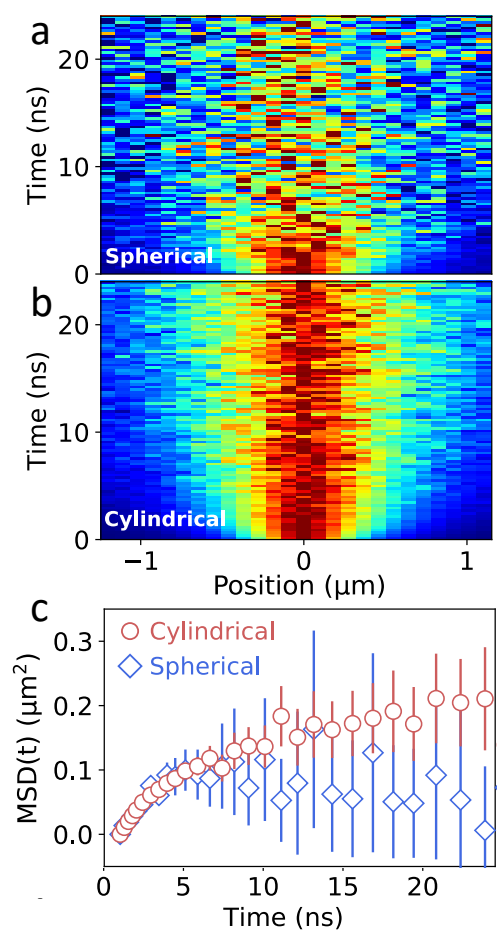

**Figure 3.** Diffusion maps of a perovskite sample for the spherical (a) and cylindrical (b) lens setups. (c) MSD extracted from (a) and (b) for the cylindrical (red circles) and the spherical (blue diamonds) setups.

In a conventional TPLM experiment with spherical lenses, the MSD is extracted by measuring along the central cross-section of the exciton population and tracking its change in variance $\sigma^2(t) - \sigma^2(0)$ for each time slice by fitting the experimental result with the expected intensity profile – normally a Gaussian or a Voigt distribution. Using a Gaussian or a Voigt distribution as a fit function for the cross section is possible in a conventional TM experiment with spherical lenses because the detector size is much smaller than the magnified image of the carrier population (20 vs. 200 µm in the example above) and can therefore be well approximated as a point detector.[10] Crucially though, because of the smaller magnification in the orthogonal direction in the cylindrical lens setup, the approximation of a point detector no longer holds. In fact, due to the small magnification in the orthogonal direction (x-direction in Figure 1 and Figure 2b) the detector no longer collects a one-dimensional slice of the distribution, but a larger collection area, which can be represented by the integral: $h(y) = \int_{-d/2}^{d/2} f(x,y)dx$, where $f(x,y)$ is the carrier distribution and d is the projected and demagnified detector size in the x direction (e.g. for a 20 µm APD and a cylindrical setup with 20× magnification: $d$ = 20 µm / 20 = 1 µm). This integral represents the expected intensity distribution to be measured for a cylindrical lens setup, and hence $h(y)$ has to be used to fit the experimental results. Interestingly, for a Gaussian exciton population, the integral is again Gaussian just like the simple cross section ($h(y) \propto f(x, y = const.)$). However, this proportionality is specific to the Gaussian fit and does not hold for other types of fit functions to the distribution. A number of Transient Microscopy studies have shown that simple Gaussian functions often fail to capture the tails of the emission profile in thin films.[2,6,14] For a more accurate determination of the MSD(t), the tails of emission profiles are often best described by Voigt functions, which are convolutions of a Lorentzian with a Gaussian.[2,6] To accurately determine the change in variance $\sigma(t)^2 - \sigma(t)^2$ of the Voigt function in the case of the definite integration, we introduce a numerical integration $h(y)$ of a 2D Voigt along the x-axis in the APD scan area. With the resulting integrated function, we fit the experimental TPLM data, allowing us to calculate the MSD(*t*) dependence using $\mathrm{MSD}(t) = \sigma(t)^2 - \sigma(0)^2$.

**Figure 3**c shows a comparison of the resulting MSD as a function of time for both the spherical and cylindrical lens setup. The error bars represent the uncertainty (one standard deviation) in the fitting procedure for MSD(*t*), in both cases increasing significantly at later times when the fluorescent signal is diminished due to the exponential decay of the excitons in the system. Importantly though, the error bars for the cylindrical lens setup are dramatically reduced as compared to the spherical lens, as expected from the improved signal collection (see **Figure S3** and **S4**). Consequently, for equal acquisition times, more precise information about the MSD can be obtained using the cylindrical lens setup, particularly at later times. Late time dynamics of the MSD can provide valuable information about anomalous transport regimes and the corresponding local energy landscape of the material.[13]

To determine the diffusion length, we use a biexponential fit of the fluorescence lifetime decay dynamics (see **Figure S5**) to determine the fraction of surviving excitons as a function of time (see **Figure S5**). The diffusion length $L_D$ is then defined as the net displacement in one dimension achieved by 37% (i.e., 1/e) of the exciton population. The improved SNR in both the spatial as well as the temporal decay dynamics using the cylindrical lens setup improves the precision with which the diffusion length can be determined for equal acquisition times, yielding $L_D = 174 \pm 11$ nm for the spherical and $L_D = 184 \pm 4$ nm for the cylindrical lens setup.

The ability to resolve the displacement of excitons is primarily determined by the SNR in the experiment. While excitons in (PEA)$_2$PbI$_4$ diffuse across several hundreds of nm during their lifetime, displacements in other systems are often much smaller. An example of such a system are colloidal quantum-dot thin films in which transport of excitons relies on Förster-mediated energy transfer between individual quantum dots in the lattice. **Figure 4** shows the comparison of the MSD as a function of time between the spherical and cylindrical lens setup. Comparing **Figure 4**a and b, a clear improvement in the SNR is observed when using the cylindrical lens setup. Importantly, even though the broadening of the population in both diffusion maps is much less pronounced than for the perovskite materials, a clear broadening can be observed in the corresponding MSD curves (see **Figure 4**c). In the case of the QDs, the MSD shows a clear sub-linear behaviour, indicative of subdiffusive transport (see **Figure 4**c). Subdiffusion of carriers in QD solids is commonly observed and is caused by polydispersity in the QD ensemble leading to a disordered energy landscape.[2] To quantify the subdiffusive behaviour, we use a power law (MSD($t$) = $2Dt^{\alpha}$). From the fits we obtain $D = 0.003 \pm 0.001$ cm$^2$/s and $D = 0.003 \pm 0.002$ cm$^2$/s in the cylindrical and spherical lenses, respectively. Corresponding diffusion lengths that are obtained from these measurements are $L_D = 41 \pm 6$ nm for the spherical and $L_D = 40 \pm 5$ nm for the cylindrical lens setup. With the cylindrical lens setup, a single measurement provides higher accuracy for both the diffusion length and the diffusivity (see **Figure S4** and **S6**).

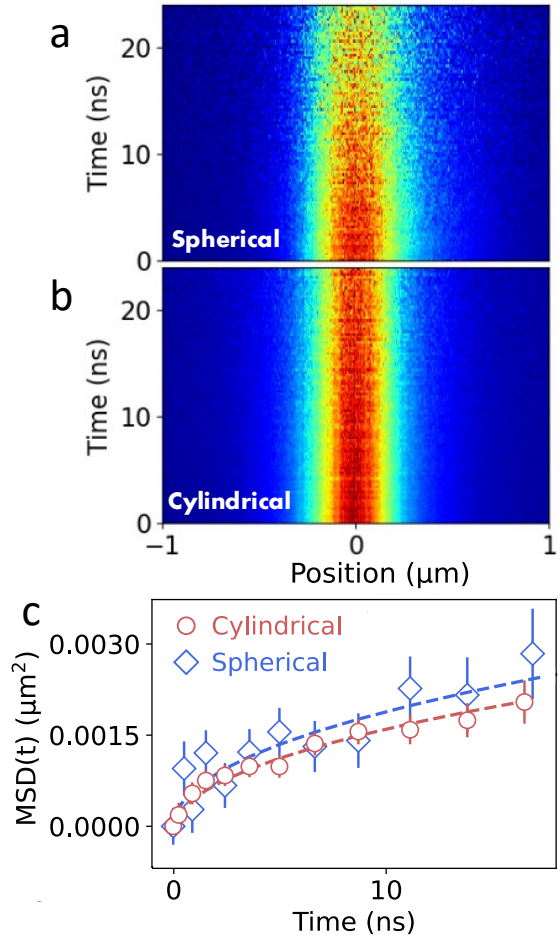

**Figure 4.** Diffusion maps of a QD sample for the spherical (a) and cylindrical (b) lens setups. (c) MSD for the cylindrical (red circles) and the spherical (blue diamonds) lenses. Dashed lines are fits to a power law.

## 3. Conclusion

We have demonstrated a simple method to improve the data acquisition efficiency in Transient Microscopy measurements for materials with isotropic diffusion in the x-y plane using cylindrical lenses. The intensity enhancement for the specific examples in this work is around 10-fold, reducing the needed acquisition time for measurements with equal SNR by the same amount. Importantly, further improvements in the intensity enhancement can be achieved using larger focal length ratios for the two cylindrical lenses, as well as using higher quality lenses. In the presented work, the combination of the two cylindrical lenses ($f_1 = 400$ mm, $f_2 = 20$ mm) were chosen for practical reasons to fit the existing path-length between microscope and detector. Naturally, if chosen freely, larger focal length ratios will yield larger compression along the x-axis and further improved intensity enhancement. Crucially though, the current setup employs a $f_2 = 20$ mm focal length lens with a diameter of 25 mm as the short focal length cylindrical lens, resulting in a small f-number and relatively poor quality of the focus. This limitation is clear when comparing the theoretical demagnification (400 / 20 = 20) and the experimentally obtained demagnification of around 12. Larger f-numbers on both cylindrical lenses would further improve the experimental intensity enhancement and the improvement in the signal collection, although at the cost of a longer optical path of the total setup. Importantly, the intensity enhancement for the cylindrical lens setup also depends on the diffusion length of the carriers in the specific material under study. For materials with long diffusion lengths, larger enhancements can be obtained as the carriers diffuse farther away from the axis of measurement in the conventional spherical lens setup. To simulate the maximum enhancement for different time-integrated population distributions and different magnification ratios for a given detector size (here taken as 20 μm) is displayed in **Figure 5**. Enhancements of up to 25 times are achievable for example for a magnification ratio of 30 and a relatively long diffusion length with a full-width-at-half-maximum (FWHM) of 1.2 μm. Simulations of the dependence of the signal enhancement on more general system parameters are shown in Figure S7.

Finally, it is worth mentioning that the use of cylindrical lenses can benefit to other Transient Microscopy techniques beyond TPLM, including for example Transient Absorption Microscopy or Transient Scattering Microscopy (i.e. stroboSCAT).[15] The improved SNR in these types of measurements will help to reduce data acquisition times and sample degradation, and may help resolve details of the transport dynamics at longer times when the decay of the carrier population reduces the signal strength.

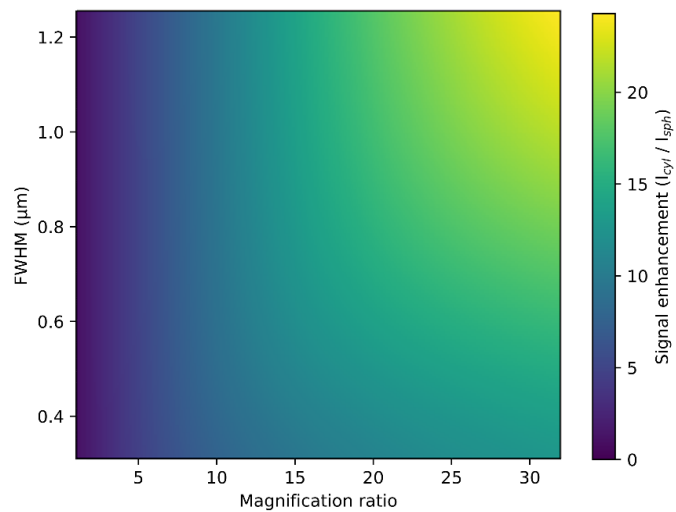

**Figure 5.** Simulations of the dependence of the signal enhancement ($I_{cyl}/I_{sph}$) on the magnification ratio and the time integrated FWHM of the energy carrier population for a detector size of 20 μm. A signal enhancement up to ~25 is predicted for magnification ratios of 30 and FWHM of 1.2 μm.

**Acknowledgements**

We acknowledge the support from the "(MAD2D-CM)-UAM" project funded by Comunidad de Madrid, by the Recovery, Transformation and Resilience Plan, and by NextGenerationEU from the European Union, as well as from the Spanish Ministry of Science and Innovation under grant agreement TED2021-131018B-C21.

**References**

[1] G. M. Akselrod, P. B. Deotare, N. J. Thompson, J. Lee, W. A. Tisdale, M. A. Baldo, V. M. Menon, V. Bulovic, *Nature Communications* **2014**, *5*, 3646.
[2] G. M. Akselrod, F. Prins, L. V. Poulikakos, E. M. Y. Lee, M. C. Weidman, A. J. Mork, A. P. Willard, V. Bulović, W. A. Tisdale, V. Bulovic, W. A. Tisdale, *Nano Letters* **2014**, *14*, 3556.
[3] E. Penzo, A. Loiudice, E. S. Barnard, N. J. Borys, M. J. Jurow, M. Lorenzon, I. Rajzbaum, E. K. Wong, Y. Liu, A. M. Schwartzberg, S. Cabrini, S. Whitelam, R. Buonsanti, A. Weber-Bargioni, *ACS Nano* **2020**, *14*, 6999.
[4] A. J. Goodman, D. H. Lien, G. H. Ahn, L. L. Spiegel, M. Amani, A. P. Willard, A. Javey, W. A. Tisdale, *Journal of Physical Chemistry C* **2020**, *124*, 12175.
[5] M. Kulig, J. Zipfel, P. Nagler, S. Blanter, C. Schüller, T. Korn, N. Paradiso, M. M. Glazov, A. Chernikov, *Physical Review Letters* **2018**, *120*, 207401.
[6] M. Seitz, A. J. Magdaleno, N. Alcázar-Cano, M. Meléndez, T. J. Lubbers, S. W. Walraven, S. Pakdel, E. Prada, R. Delgado-Buscalioni, F. Prins, *Nature Communications* **2020**, *11*, 1.
[7] J. D. Ziegler, J. Zipfel, B. Meisinger, M. Menahem, X. Zhu, T. Taniguchi, K. Watanabe, O. Yaffe, D. A. Egger, A. Chernikov, *Nano Letters* **2020**, *20*, 6674.
[8] W. Li, M. Shao Ran Huang, S. K. Yadavalli, J. David Lizarazo Ferro, Y. Zhou, A. Zaslavsky, N. P. Padture, R. Zia, *ACS Photonics* **2019**, *6*, 2375.
[9] M. I. Saidaminov, K. Williams, M. Wei, A. Johnston, R. Quintero-Bermudez, M. Vafaie, J. M. Pina, A. H. Proppe, Y. Hou, G. Walters, S. O. Kelley, W. A. Tisdale, E. H. Sargent, *Nature Materials* **2020**, *19*, 412.
[10] N. S. Ginsberg, W. A. Tisdale, *Annual Review of Physical Chemistry* **2020**, *71*, 1.
[11] S. Deng, E. Shi, L. Yuan, L. Jin, L. Dou, L. Huang, *Nature Communications* **2020**, *11*, 1.
[12] A. J. Magdaleno, M. Seitz, M. Frising, A. Herranz de la Cruz, A. I. Fernández-Domínguez, F. Prins, *Materials Horizons* **2021**, *8*, 639.
[13] M. Seitz, M. Meléndez, N. Alcázar-Cano, D. N. Congreve, R. Delgado-Buscalioni, F. Prins, *Advanced Optical Materials* **2021**, *2021*, 2001875.
[14] C. L. Hickey, E. M. Grumstrup, *Journal of Physical Chemistry C* **2020**, *124*, 14016.
[15] M. Delor, H. L. Weaver, Q. Q. Yu, N. S. Ginsberg, *Nature Materials* **2020**, *19*, 56.

# Supporting information

## Experimental methods

**2D perovskite preparation**

Perovskite solutions were made in a $N_2$-filled glovebox following the recipes given in refs.[1–3] In short, n = 1 phenethylammonium lead iodine $(PEA)_2PbI_4$ solutions were prepared by mixing stoichiometric ratios of the precursor salts. Phenethylammonium iodide (PEAI) (Sigma Aldrich, 805904-25G) and lead(II) iodide ($PbI_2$) (Sigma Aldrich, 211168-50G) were mixed with a stoichiometric ratio of 2:1 and dissolved in anhydrous n,n-dimethylformamide (DMF, Sigma Aldrich, 227056-1L). The solution was heated to 70°C while stirring until all the precursors were completely dissolved. The resulting solution was kept at 70°C and the solvent was left to evaporate until reaching saturation of the solution.

Perovskite crystals (hundreds of micron lateral sizes) were obtained using drop casting of the saturated perovskite solution on top of a microscope slide and heating to 50°C for 2-3 hours until drying. The perovskite crystals of the thin film were mechanically exfoliated using the Scotch tape (Scotch Magic) method. After several exfoliation steps, the crystals were transferred on a glass slide and were subsequently studied through the glass slide with a ×100 oil immersion objective (Nikon CFI Plan Fluor, NA = 1.3).

**CdSe QD samples**

CdSe/ZnS (Sigma Aldrich, 790192) thin films are made to carry out the measurements. The surface functionalization of octadecyltrichlorosilane (OTS, Sigma Aldrich, 104817-25G) based self-assembled monolayers explained by Prins, et al[4] is applied to increase the hydrophobicity of the glass substrates (24x24x0.16 mm, Labbox, COVN-024-200). It is needed to get smooth and densely packed QD films for the diffusion measurements. For safety reasons, the surface functionalization process of Prins, et al[4] is slightly modified, using a plasma treatment instead of the piranha cleaning for the generation of the oxygen radicals in the glass surface. The plasma cleaning was applied during 1 min, the chosen gas to create the plasma was air.

CdSe/ZnS solutions of ~40 mg/mL[5] are prepared using a hexane:octane ratio of 9:1. Subsequently, the QD solutions are drop casted on the functionalized glass substrates.

**Emission profile measurements**

The materials were excited with a 405 nm laser (PicoQuant LDH-D-C-405, PDL 800-D), which was focused down to a near diffraction limited spot. The laser spot images were acquired using an EMCCD camera coupled to a spectrograph (Princeton Instruments, SpectraPro HRS-300, ProEM HS 1024BX3) with a x430 magnification for the spherical and x430 (y-axis), x32 (x-axis) for the cylindrical lenses configuration, i.e. 100x oil immersion objective with the additional magnification of the respective lenses. Results are shown in **Figure 2**.

**Exciton propagation length**

Fluorescence lifetime measurements were performed using a laser diode of λ = 405 nm (PicoQuant LDH-D-C-405, PDL 800-D, Pico-Harp 300) and an avalanche photodiode (APD, Micro Photon Devices PDM). The repetition rate was 40 MHz and the peak fluence per pulse was 50 nJ cm$^{-2}$. **Figure S5** shows the photoluminescence lifetime traces of the 2D perovskites and CdSe QDs and the fit parameters of a bi-exponential fit to the data. The fit was used together with the experimentally obtained time-dependent MSD to extract the total number of surviving excitons for a given time t as presented in **Figure S6**. The exciton propagation length is defined as the net distance at which 1/e of the free excitons remain. The total number of surviving excitons at time t is given by:

$$\text{surviving excitons (t)} = \frac{\int_t^\infty w_1 e^{-\frac{t}{\tau_1}} + w_2 e^{-\frac{t}{\tau_2}} dt}{\int_0^\infty w_1 e^{-\frac{t}{\tau_1}} + w_2 e^{-\frac{t}{\tau_2}} dt}$$

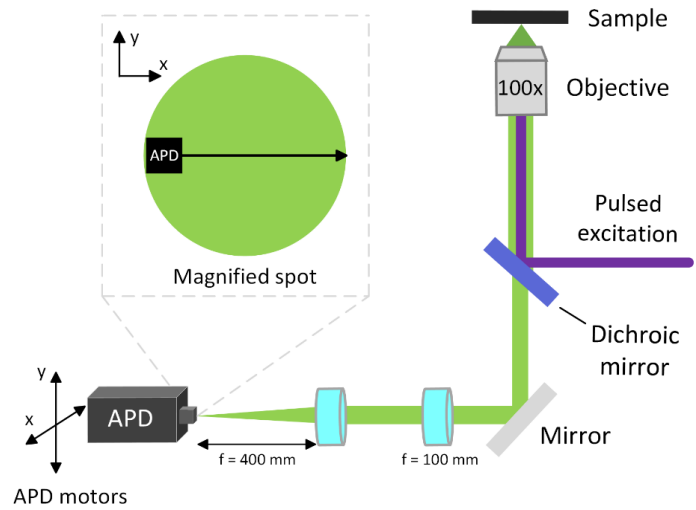

Figure S1. Conventional TPLM with spherical lenses. The sample is excited with a laser pulse. The photoluminescence spot is visualized with a x100 magnification objective. The PL light is filtered with a dichroic mirror and sent through a lens (with 100 mm focal length) and a spherical lens with a long focal length (f = 400 nm) to obtain a final magnification of around x400.

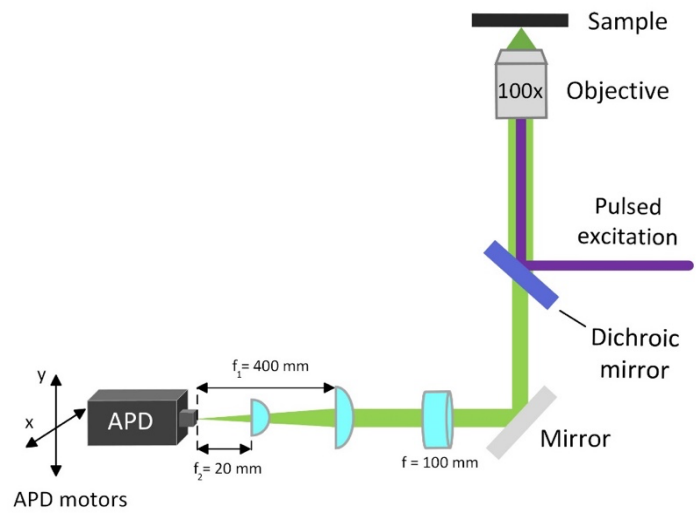

Figure S2. Optical set-up configuration with cylindrical lenses. The sample is excited with a laser pulse. The photoluminescence spot is visualized with a x100 magnification objective. The PL light is filtered with a dichroic mirror and sent through a lens (with 100 mm focal length) and a set of two cylindrical lenses. The last two project with high magnification the y-axis of the PL spot while keeping a smaller magnification on the x-axis.

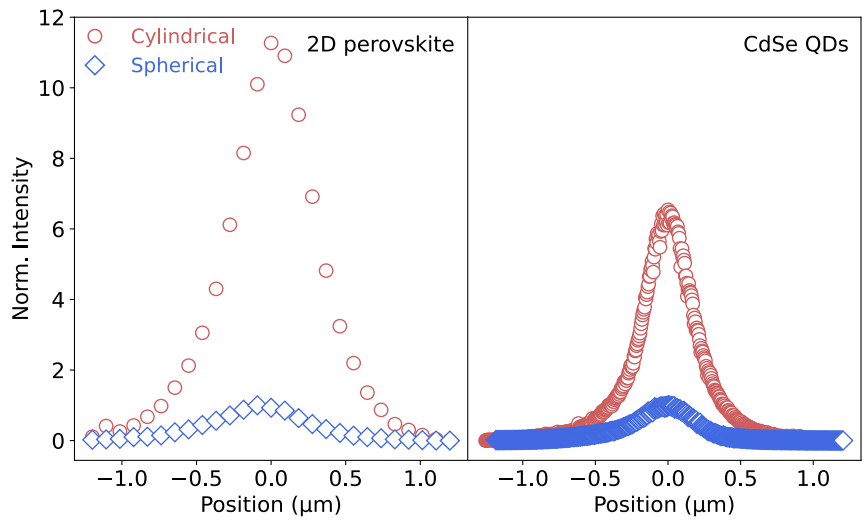

Figure S3. TPLM intensities obtained along the raster scan of the APD across the PL spot. The intensities captured with the cylindrical lenses are depicted in red and with the spherical lens in blue. For a better comparison the values are normalized to the maximum intensity measured with the spherical lens configuration.

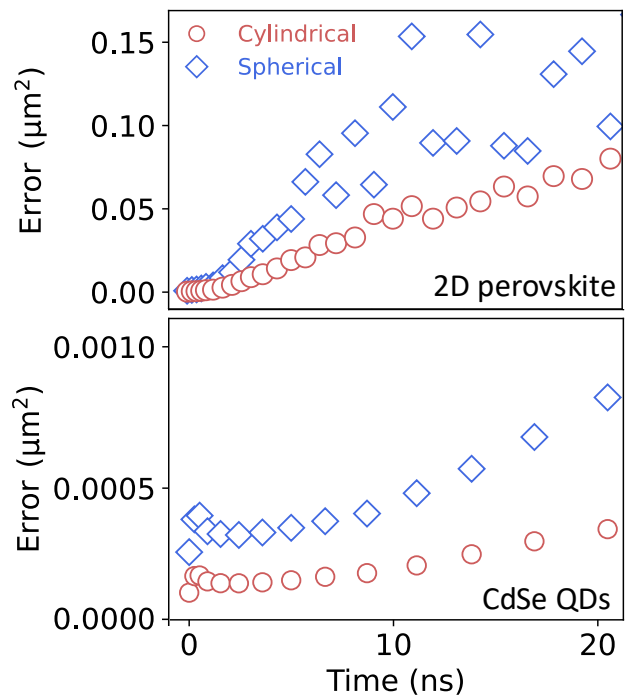

Figure S4. Error of the fits to the spatial distribution at each time-slice for both the spherical (blue diamonds) and cylindrical (red circles) lens setups associated with the diffusion measurements shown in **Figure 3** (2D perovskite) and **4** (CdSe QDs), respectively.

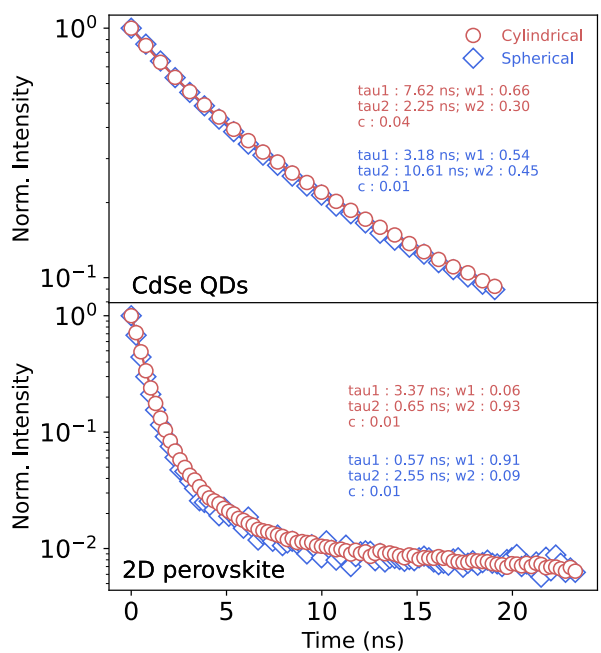

Figure S5. Photoluminescence lifetime decays measured with the cylindrical (red) and the spherical (blue) lenses fitted with a biexponential fit for both the CdSe QDs and 2D perovskites. Fitting parameters are given in the figure.

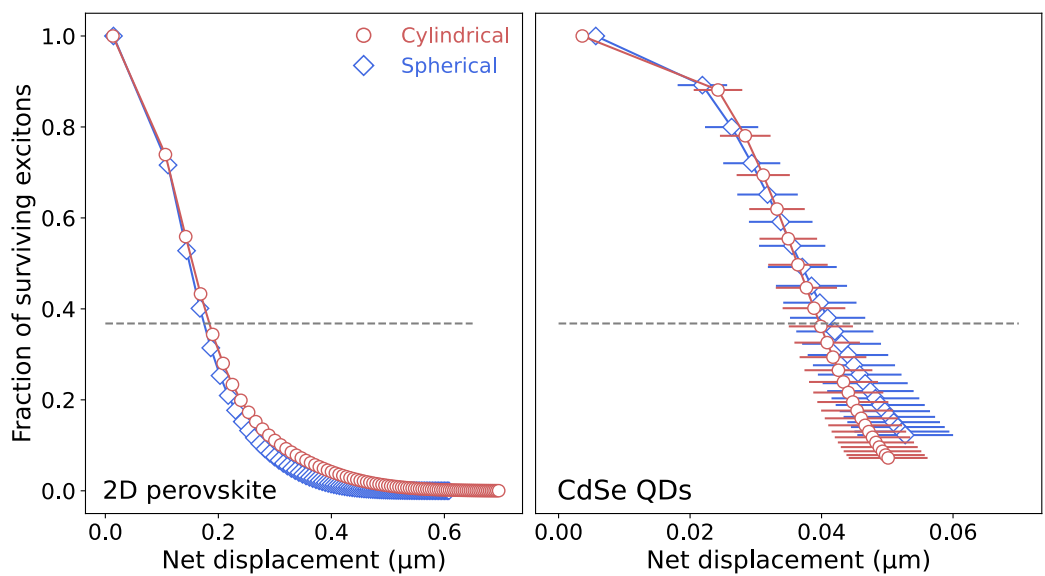

Figure S6. Fraction of surviving excitons (extracted from lifetime data in Supplementary **Figure S5**) as a function of net spatial displacement $\sqrt{MSD(t)}$ for the cylindrical (red) spherical (blue) lenses for both the 2D perovskites and CdSe QDs. Reported errors represent the uncertainty in the fitting procedure for σ(t)². Dashed grey line represents that 1/e of the excitons have survived.

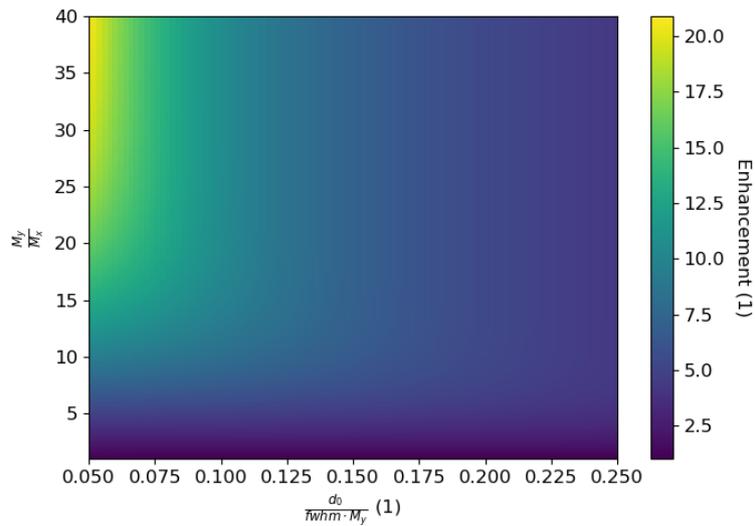

Figure S7 Simulations of the dependence of the signal enhancement ($I_{cyl}/I_{sph}$) on more general system parameters: $d_0$ being the detector size, fwhm the population size, and $M_y$ and $M_x$ magnifications in y and x directions, respectively.

## References

[1] S. T. Ha, C. Shen, J. Zhang and Q. Xiong, *Nat. Photonics*, 2016, **10**, 115–121.
[2] M. Seitz, P. Gant, A. Castellanos-Gomez and F. Prins, *Nanomaterials*, 2019, **9**(8), 1120
[3] M. Seitz, A. J. Magdaleno, N. Alcázar-Cano, M. Meléndez, T. J. Lubbers, S. W. Walraven, S. Pakdel, E. Prada, R. Delgado-Buscalioni and F. Prins, *Nat. Commun.*, 2020, **11**, 1–8.
[4] F. Prins, D. K. Kim, J. Cui, E. De Leo, L. L. Spiegel, K. M. Mcpeak and D. J. Norris, *Nano Lett.*, 2017, 17, 3, 1319–1325
[5] G. M. Akselrod, F. Prins, L. V. Poulikakos, E. M. Y. Lee, M. C. Weidman, A. J. Mork, A. P. Willard, V. Bulović and W. A. Tisdale, *Nano Lett.*, 2014, **14**, 3556–3562.